\documentclass[aps,prl,twocolumn,showpacs,preprintnumbers,superscriptaddress]{revtex4}

\usepackage{graphicx}
\usepackage{dcolumn}
\usepackage{bm}
\newcommand{\CFA}{\rm CaFe_{4}As_{3}}

\begin{document}

\title{Thermal and electrical transport in the spin density wave antiferromagnet  CaFe$_{4}$As$_{3}$}

\author{M. S. Kim}
\affiliation{Condensed Matter Physics and Materials Science
Department, Brookhaven National Laboratory, Upton, New York
11973-5000, USA}\affiliation{Department of Physics and Astronomy,
Stony Brook University, Stony Brook, New York 11794-3800, USA}
\author{Z. P. Yin}
\affiliation{Department of Physics and Astronomy, Stony Brook
University, Stony Brook, New York 11794-3800, USA}
\affiliation{Department of Physics and Astronomy, Rutgers
University, Piscataway, NJ 08854.}
\author{L. L. Zhao}
\affiliation{Department of Physics and Astronomy, Rice
University, Houston, Texas 77005, USA}
\author{E. Morosan}
\affiliation{Department of Physics and Astronomy, Rice
University, Houston, Texas 77005, USA}
\author{G. Kotliar}
\affiliation{Department of Physics and Astronomy, Rutgers
University, Piscataway, NJ 08854.}
\author{M. C. Aronson}
\affiliation{Condensed Matter Physics and Materials Science
Department, Brookhaven National Laboratory, Upton, New York
11973-5000, USA}\affiliation{Department of Physics and Astronomy,
Stony Brook University, Stony Brook, New York 11794-3800, USA}

\date{\today}

\begin{abstract}
We present here measurements of the thermopower, thermal
conductivity, and electrical resistivity of the newly reported
compound CaFe$_{4}$As$_{3}$. Evidence is presented from specific
heat and electrical resistivity measurements that a substantial
fraction of the Fermi surface survives the onset of spin density
wave (SDW) order at the N\'eel temperature $T_{\rm N}$=88 K, and
its subsequent commensurate lockin transition at $T_{2}$=26.4 K.
The specific heat below $T_{2}$ consists of a normal metallic
component from the ungapped parts of the Fermi surface, and a
Bardeen-Cooper- Schrieffer (BCS) component that represents the
SDW gapping of the Fermi surface. A large Kadowaki-Woods ratio is
found at low temperatures, showing that the ground state of
CaFe$_{4}$As$_{3}$ is a strongly interacting Fermi liquid.  The
thermal conductivity $\kappa$ of CaFe$_{4}$As$_{3}$ is an order
of magnitude smaller than those of conventional metals at all
temperatures, due to a strong phonon scattering. The
thermoelectric power $S$ displays a sign change from positive to
negative indicating that a partial gap forms at the Fermi level
with the onset of commensurate spin density wave order at
$T_{2}=26.4$ K.  The small value of the thermopower $S$ and the
enhancements of the resistivity due to gap formation and strong
quasiparticle interactions offset the low value of the thermal
conductivity $\kappa$, yielding only a modest value for the
thermoelectric figure of merit $\mathcal{Z}\leq 5\times10^{-6}
K^{-1}$ in CaFe$_{4}$As$_{3}$. The results of ab initio
electronic structure calculations are reported, confirming that
the sign change in the thermopower at $T_{2}$ is reflected by a
sign change in the slope of the density of states at the Fermi
level. Values for the quasiparticle renormalization $Z$ are
derived from measurements of the specific heat and thermopower,
indicating that as $T\rightarrow0$, $\CFA$ is among the most
strongly correlated of the known Fe-based pnictide and
chalcogenide systems.

\end{abstract}

\pacs{72.15.Eb, 72.15.Jf, 75.30.Fv}

\maketitle

\section{Introduction}

Although thermoelectric materials have received enduring
attention over the past half century for their ability to convert
thermal energy to electrical energy and vice versa, the growing
need to develop new materials with enhanced properties for
applications has led to a renewed interest in recent
years.~\cite{nolas0} The potential performance of a given
thermoelectric material is expressed in terms of the
thermoelectric figure of merit $\mathcal{Z}=S^2/\rho\kappa$,
where $S$ is the thermoelectric power, $\rho$ is the electrical
resistivity, and $\kappa$ is the thermal conductivity. High
thermoelectric power must be combined with low thermal
conductivity and low electrical resistivity to optimize the
figure of merit. Large values of $\mathcal{Z}$ have been observed
in the filled skutterudites $AT_{4}$Sb$_{12}$ ($A$ = rare earths
or alkaline earth metals, $T$ = transition metals), which combine
large values of the thermopower $S$, which can be as large as
10-100 $\mu$V/K, with low thermal conductivity $\kappa$, as small
as a few W/K-m.~\cite{sales0,takabatake} The low $\kappa$ of
these systems reflects an unusually small phonon contribution,
originating with enhanced scattering of heat-carrying phonons
from the low energy rattling and tunneling modes of filler
``$A$'' atoms in the $T_{4}$Sb$_{12}$ cage-like structures
characteristic for these compounds.~\cite{sales,goto}
Consequently, it is potentially of great interest to identify and
explore other classes of correlated materials that similarly form
in open and cage-like structures.

Recently, a new Fe-As system, CaFe$_{4}$As$_{3}$, has been
reported,~\cite{todorov,zhao} that crystallizes in orthorhombic
structure (space group $Pnma$). In this structure, the Ca atoms
are confined inside nearly rectangular tunnels, oriented along the
$b$ axis, whose walls are constructed from a network of Fe-As
tetrahedra which are similar to the FeAs planes in the layered
iron-pnictide superconductors. The combination of confined Ca
atoms and the open structure of the FeAs framework in
CaFe$_{4}$As$_{3}$  is reminiscent of the cages in the filled
skutterudites $R$Fe$_{4}$As$_{12}$ ($R$= Ce and
Pr).~\cite{watch,sayles}, and it is possible that here too a low
thermal conductivity might be realized. We note as well that
large thermopowers are generally found in metals where strong
correlations lead to enhanced densities of states near the Fermi
level. CaFe$_{4}$As$_{3}$ undergoes an antiferromagnetic
transition at $T_{\rm N}= 88$ K, which has been classified as an
incommensurate spin density wave (SDW).~\cite{nambu} The SDW
becomes commensurate with the underlying lattice below a second
transition that occurs at $T_{2}=26.4$ K. However, the SDW does
not appear to gap the entire Fermi surface, since metallic Fermi
liquid behavior is found at the lowest temperatures, where the
electrical resistivity $\rho(T)=\rho_{0}+{\rm A}T^{2}$ and the
electronic specific heat $C_{\rm el}=\gamma T$. Intriguingly, the
Kadowaki-Woods ratio ${\rm A}/\gamma^{2}$ is very large,
approaching a value of $55\times10^{-5}$ $\mu\Omega$cm mol$^{2}$
K$^{2}$ mJ$^{-2}$.~\cite{zhao} This suggests that the
quasiparticles of the Fermi liquid have substantial interactions,
at a level that is comparable to those realized in the heavy
fermion compounds. The possibility of a small thermal
conductivity, derived from the confinement of the Ca atoms and
the possibility of strong enhancement of the density of states at
low temperatures, suggested by the large Kadowaki-Woods ratio,
make CaFe$_{4}$As$_{3}$ a promising candidate for a large
thermoelectric figure of merit. Accordingly, we present here
measurements of the thermal and electrical  transport properties
of single crystals of CaFe$_{4}$As$_{3}$, which together with ab
initio calculations of the electronic structure, probe not only
the character of the quasiparticles forming the ground state, but
also the impact of the SDW transitions on the electronic
structure.

\section{Experimental Details}

Single crystals of CaFe$_{4}$As$_{3}$ were grown from a Sn flux,
forming in a rodlike morphology with the crystallographic $b$
axis along the rod axis. Details of the sample preparation are
described in Ref.~\onlinecite{zhao}. Using the thermal transport
option (TTO) of a Quantum Design Physical Property Measurement
System (PPMS) over the range of temperatures with 2 K $<T<$ 300
K, the thermal conductivity $\kappa(T)$, thermoelectric power
$S(T)$, and resistivity $\rho(T)$ were simultaneously measured on
a $0.5\times0.5\times3$ mm$^3$ sample using a two-probe method,
where the temperature gradient across the sample was always less
than 3\% of the background temperature during the measurement and
an AC current of 1 mA with frequency of 17 Hz was applied for the
resistivity measurement. Specific heat measurements were also
carried out using the thermal relaxation technique implemented on
the PPMS for 5 K $<T<$ 150 K.


\section{Experimental Results and Discussion}

\begin{figure}
\includegraphics[width=7.0cm]{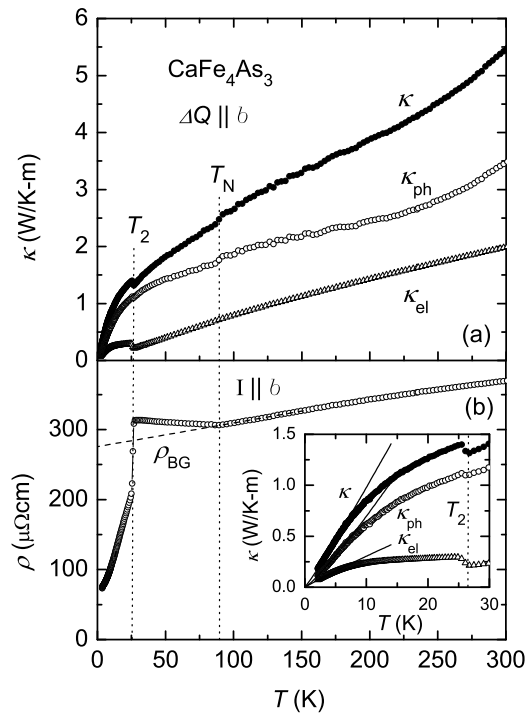}
\caption{(a) Temperature dependencies of the measured thermal
conductivity $\kappa$ along the $b$ axis, the electron
contribution of the thermal conductivity $\kappa_{\rm el}$
estimated from the Wiedemann-Franz law, and the phonon
contribution of the thermal conductivity $\kappa_{\rm
ph}(=\kappa-\kappa_{\rm el})$, see text. (b) Temperature
dependence of the resistivity $\rho$ along the $b$ axis (open
circles). The dashed line indicates $\rho_{\rm BG}$ which is
determined from a linear extrapolation of the data above $T_{\rm
N}$. The inset shows $\kappa$, $\kappa_{\rm el}$, and
$\kappa_{\rm ph}$ below 30 K. \label{fig1}}
\end{figure}

The filled circles in Fig.~\ref{fig1}(a) show the temperature
dependence of the thermal conductivity $\kappa$, measured with
the heat flow along the $b$ axis of a single crystal of
CaFe$_{4}$As$_{3}$. The temperature dependence of the electrical
resistivity $\rho(T)$ is presented in Fig.~\ref{fig1}(b), showing
an overall metallic character, with a sharp drop at $T_{2}$ and a
slope discontinuity at $T_{\rm N}$. The electronic contribution
to the thermal conductivity $\kappa_{\rm el}$ was determined
using the Wiedemann-Franz law $\kappa_{\rm el}=L_{0}T/\rho$,
where $L_{0}$ is the Sommerfeld value $2.45\times10^{-8}$
W$\Omega$/K$^{2}$ for the Lorenz ratio $L=\rho\kappa/T$. The
phonon contribution $\kappa_{\rm ph}$ was subsequently determined
as the difference between $\kappa$ and $\kappa_{\rm el}$.
$\kappa$, $\kappa_{\rm el}$, and $\kappa_{\rm ph}$ are compared
in Fig.~\ref{fig1}(a).  We note that values for $\kappa$ above
$\sim$ 200K are likely to be $\sim$10-30$\%$ higher than the
actual values, due to systematic discrepancies in the PPMS/TTO
radiation corrections. ~\cite{rudajevova,sebek}.

The overall thermal conductivity $\kappa$ of CaFe$_{4}$As$_{3}$
is an order of magnitude smaller at all temperatures than the
values found in conventional metals, which are typically
$\approx$10-30 W/K-m. However $\kappa$ is comparable to the values
measured in the filled skutterudites
$AT_{4}X_{12}$.~\cite{watch,sayles,takabatake,narazu} For
example, $\kappa$ of $R$Fe$_{4}$As$_{12}$ ($R=$ Ce and Pr) is no
more than $3\sim4$ W/K-m at all temperatures,~\cite{watch,sayles}
comparable to the  values found in CaFe$_{4}$As$_{3}$ above
$T_{\rm N}$ (Fig.~\ref{fig1}(a)), and as much as a full order of
magnitude larger than at the lowest temperatures. By analogy to
the filled skutterudites, the characteristically small value of
$\kappa$ in CaFe$_{4}$As$_{3}$ can be attributed to its cage-like
structure, described above.~\cite{todorov,zhao} The equivalent
isotropic atomic displacement parameter (ADP) of the Ca atoms is
more than 40\% larger than the averaged equivalent isotropic ADP
of the Fe and As atoms in CaFe$_{4}$As$_{3}$,~\cite{todorov} and
is comparable to the amplitudes of motion realized in the
skutterudites, where the ADPs of the filler atoms are $\approx$
30\% larger than those of the other atoms.~\cite{bauer} If the low
thermal conductivity observed in CaFe$_{4}$As$_{3}$ arises from
the rattling of Ca atoms in the open tunnel-like structure, then
the phonon mean free path should be comparable to either the 6
\AA~cage dimension, or to the 3.7 \AA~nearest Ca-Ca separation
distance. We have estimated the phonon mean free path $d$ using
the expression $\kappa_{\rm ph}=\frac{1}{3}C_{v}v_{s}d$, where
$C_{v}$ at 300 K is 189 J/mol K, a Debye temperature $\theta_{\rm
D}$ is taken from the ADP of CaFe$_{4}$As$_{3}$ as 328 K, and the
averaged sound velocity $v_{s}$ is estimated from the Debye model
to be 2784 m/s.~\cite{sales2} The total mean free path $d\sim12$
\AA, which is substantially larger than either the dimension of
the tunnel-like structure in  CaFe$_{4}$As$_{3}$ or the spacing of
the Ca ions contained in the cage. We conclude that the small
thermal conductivity likely doesn't originate with a rattling
mode, but instead with strong phonon scattering due to the
complicated crystal structure and 32 atom unit cell, with
multiple Fe and As site symmetries.

Further evidence for the strong phonon scattering comes from the
temperature dependence of $\kappa$, which displays a broad
shoulder centered around 30 K in CaFe$_{4}$As$_{3}$. Ordinarily,
a broad maximum in $\kappa_{\rm ph}$ is expected at $T \approx
\theta_{\rm D}/10$, resulting from the crossover between
phonon-boundary and/or phonon-point defect scattering at low
temperatures ($T\ll\theta_{\rm D}/10$) and phonon-phonon Umklapp
scattering at high temperatures ($T\gg\theta_{\rm D}/10$). In
principle, the absence of a maximum in $\kappa(T)$ in
CaFe$_{4}$As$_{3}$ might be ascribed to particularly strong
phonon scattering from point-like defects. However, the broad
maximum is found even in polycrystalline filled skutterudites,
where defects are expected to be plentiful.~\cite{takabatake} A
shoulder in $\kappa(T)$ is found only in skutterudites with small
filler atoms,~\cite{nolas, takabatake} whose large amplitude
rattling leads to particularly strong phonon scattering. The
shoulder observed in $\kappa(T)$ in CaFe$_{4}$As$_{3}$ implies
that the phonon scattering is very strong here as well.

The temperature dependencies of the thermal conductivities
$\kappa$ and the electrical resistivity $\rho$ reveal the
presence of a SDW transition into an antiferromagnetically
ordered state below the N\'eel temperature $T_{\rm N}$=88 K.
Recent neutron diffraction measurements confirm that the 88 K
transition in CaFe$_{4}$As$_{3}$ is indeed to an incommensurate
SDW state.~\cite{nambu} Figure~\ref{fig1}(a) shows that
$\kappa_{\rm el}$ decreases approximately linearly as the
temperature is reduced, and like $\rho(T)$, displays a weak
anomaly at $T_{\rm N}$. Measurements carried out on
charge-density wave (CDW) systems K$_{0.3}$MoO$_{3}$ and
(TaSe$_{4}$)$_{2}$I,~\cite{kwok} find only a small cusp in
$\kappa_{\rm ph}$  but not in $\kappa_{\rm el}$ with the onset of
the lattice distortion associated with the CDW, just as we
observe at the SDW transition in CaFe$_{4}$As$_{3}$. The
softening of the phonons that drive CDWs, and to a lesser extent
SDWs, is thought to be responsible for the reduced values of
$\kappa_{\rm ph}$ found in the ordered state $T\leq T_{\rm N}$
(Fig.~\ref{fig1}(a)).~\cite{maki} Given the complexity of the
Fermi surface of CaFe$_{4}$As$_{3}$,~\cite{nambu} we can expect
that SDW formation will lead only to a partial gapping of the
Fermi surface, resulting in a decreased density of states and a
higher electrical resistivity in the ordered phase. This explains
the initial increase in $\rho(T)$ below $T_{\rm N}$ in
CaFe$_{4}$As$_{3}$, which is much as is found in Cr near its 311 K
transition to an incommensurate SDW state.~\cite{fawcett}
However, the remaining states at the Fermi surface ultimately
lead to the resumption of a metallic temperature dependence of
the resistivity below $T_{2}$ (Fig.~\ref{fig1}(b)), previously
reported by Zhao {\it et al}.~\cite{zhao} A second resistivity
anomaly is found at $T_{2}=26.4$ K, where the SDW becomes
commensurate.~\cite{nambu} $\kappa_{\rm el}$ increases sharply at
$T_{2}$ (inset of Fig.~\ref{fig1}(b)), but a monotonic decrease
in $\kappa$, $\kappa_{\rm el}$ and $\kappa_{\rm ph}$ is found at
the lowest temperatures. In combination with the decidedly
metallic electrical resistivity (Fig.~\ref{fig1}(b)), we conclude
that the SDW gapping of the Fermi surface is only partial, and
that the residual density of states at the Fermi level leads to
normal metallic behavior as $T\rightarrow0$.

\begin{figure}
\includegraphics[width=7.0cm]{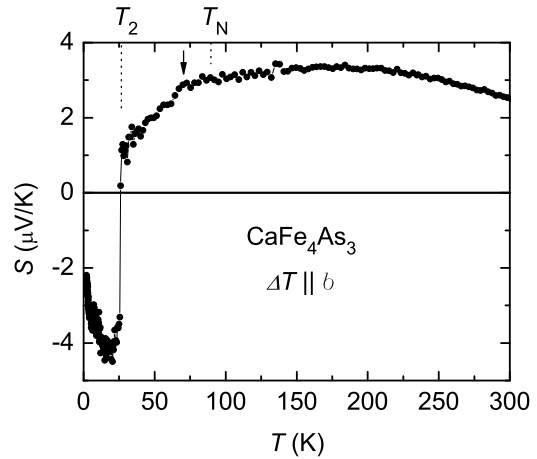}
\caption{Temperature dependence of the thermoelectric power $S$
with the heat flow along the $b$ axis. \label{fig2}}
\end{figure}

The temperature dependence of the thermoelectric power $S(T)$
depicted in Fig. \ref{fig2} also confirms the strong phonon
scattering in CaFe$_{4}$As$_{3}$. In general, the thermoelectric
power $S$ is the sum of the phonon drag and diffusion terms,
$S_{\rm g}$ and $S_{\rm d}$, respectively. Phonon-drag in metals
leads to a prominent peak in $S_{\rm g}$ at $\sim\theta_{\rm
D}/5$, which is caused by a crossover between different phonon
scattering mechanisms at higher and lower
temperatures.~\cite{blatt} The small arrow in Fig. 2 shows at
most a vanishingly small peak in $S(T)$ near 70 K $\simeq$
$\theta_{\rm D}$/5, indicating that the phonon drag peak, and
indeed $S_{\rm g}(T)$ itself, are suppressed in
CaFe$_{4}$As$_{3}$, presumably by the strong phonon scattering.
Consequently, we take $S_{\rm g}\ll S_{\rm d}$, so that $S=S_{\rm
g}+S_{\rm d}\sim S_{\rm d}$ for CaFe$_{4}$As$_{3}$.

The thermopower $S$ is surprisingly small in CaFe$_{4}$As$_{3}$,
considering the rather large magnitude of the electrical
resistivity $\rho$.  Nonetheless, the magnitudes of $S$ and
$\rho$ are comparable to those found in CaFe$_{2}$As$_{2}$, where
$\rho\simeq$ 0.4 m$\Omega$cm and $S\simeq$1-2$\mu$V/K above the
structure/magnetic transition temperature at 170
K.~\cite{matusiak} This correspondence suggests that the small
magnitude of $S$ in both compounds results from a near balance of
the electron and hole concentrations, and this result is
confirmed  for CaFe$_{2}$As$_{2}$ in  first-principle electronic
structure calculations.~\cite{ma} We will show below that our own
electronic structure calculations, carried out using DFT, support
a similar conclusion in CaFe$_{4}$As$_{3}$, and that this
particle-hole symmetry is expected in compounds with SDW
transitions.

Above $T_{2}$, $S$ is positive with a broad maximum, reaching a
value of 3.3 $\mu$V/K at $\approx$ 170 K. Only a vanishingly weak
anomaly is found at $T_{\rm N}$, while a sharp drop is found at
$T_{2}$ where $S$ changes sign from positive to negative with
decreasing temperature. $S$ is proportional to the energy
derivative of the density of states (DOS). The sign change of $S$
at $T_{2}$ indicates that a drastic change of the DOS accompanies
the commensurate SDW transition at $T_{2}$, and we will show
below that this too is consistent with the electronic structure
calculations.

\begin{figure}
\includegraphics[width=7.0cm]{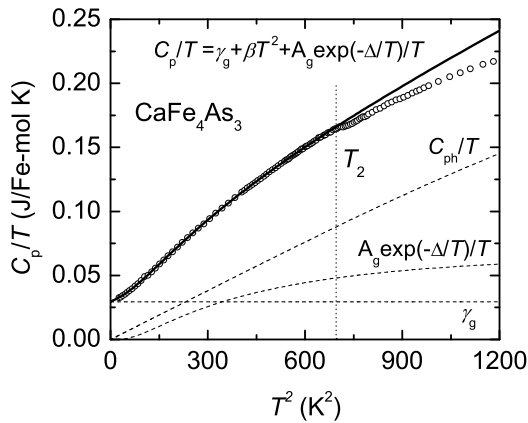}
\caption{The plot of $C_{\rm p}/T$ vs $T^{2}$ below 35 K. The
solid line indicates the fit of the expression described in the
figure, below $T_{2}$. Dashed lines indicate the phonon specific
heat divided by $T$, $C_{\rm ph}/T$ which is estimated using Debye
model with Debey temperature ($\theta_{\rm D}=312$ K) calculated
from obtained $\beta$, the Sommerfeld specific heat $\gamma_{g}$,
and BCS component divided by $T$, A$_{g}{\rm
exp}(-\Delta/T)/T$.\label{fig3}}
\end{figure}

\begin{figure}
\includegraphics[width=7.0cm]{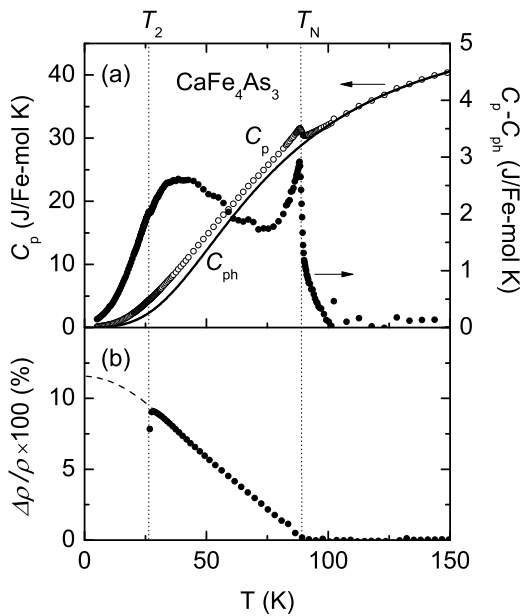}
\caption{(a) Temperature dependencies of the specific heat
$C_{\rm p}$ (open circles), the phonon contribution $C_{\rm ph}$
estimated from the Debye model (solid line), and their
difference, the electronic specific heat $C_{\rm p}-C_{\rm ph}$
(filled circles). (b) Temperature dependence of the ratio
$\Delta\rho/\rho\times100$, extrapolated below $T_{2}$ (dashed
line). \label{fig4}}
\end{figure}

Figure~\ref{fig3}(a) shows the temperature dependence of the
specific heat $C_{\rm p}/T$ measured for CaFe$_{4}$As$_{3}$ for
$5<T<150$ K. $C_{\rm p}/T$ is not linear in $T^{2}$ for $T\leq35$
K, suggesting that there may be an electronic and magnetic
component of the specific heat for $T\leq T_{2}$, in addition to
the phonon contribution. To investigate this possibility, we have
modeled $C_{\rm p}$ as the sum of a Sommerfeld term
$\gamma_{g}T$, a phonon contribution $C_{\rm ph}=\beta T^{3}$,
and a BCS term $C_{\rm BCS}={\rm A}_{g}{\rm exp}(-\Delta/T)$ that
reflects the temperature dependence in the specific heat
resulting from the formation of the SDW gap:
\begin{eqnarray*}
C_{\rm p}/T &=& C_{\rm el}/T + C_{\rm ph}/T + C_{\rm BCS}/T\\
            &=& \gamma_{g} + \beta T^{2} + {\rm A}_{g}{\rm
            exp}(-\Delta/T)/T
\end{eqnarray*}
The best fit to the data for $T\leq T_{2}$  is shown in Fig. 3,
where the three different components of the fit are compared. The
parameters of this fit are $\gamma_{g}=0.03$ J/Fe-mol
 K$^{2}$, $\beta=1.27\times10^{-4}$ J/Fe-mol K$^{3}$, A$_{g}=9.37$ J/Fe-mol
K, and $\Delta=52.80$ K. The ratio $\Delta/k_{\rm B}T_{2}=2.0$ is
very similar to the value of 2.3 found for the incommensurate SDW
in Cr,~\cite{fawcett} and is also close to the minimum value of
1.764 found in the two-band model of itinerant
antiferromagnetism.~\cite{fedders} We note that the Debye
expression using the same value of $\theta_{\rm D}=312$ K
determined from this fit also describes the specific heat above
$T_{\rm N}$ well (Fig.~\ref{fig4}(a)), where the electronic and
magnetic contributions are expected to be very small. This value
of $\theta_{\rm D}=312$ K is also in good agreement with the value
of 328 K that was determined from the ADP. The internal
consistency of these different estimates of $\theta_{\rm D}$
gives added weight to our conclusion that a broad peak at
$\sim$40 K in $C_{\rm p}-C_{\rm ph}$ is electronic and magnetic
in origin, and does not reflect anomalous features in the phonon
density of states that are not described by the Debye model. The
electronic and magnetic part of the specific heat $C_{\rm
p}-C_{\rm ph}$ displays a sharp peak at $T_{\rm N}$
(Fig.~\ref{fig4}(a)), although only a weak peak marks the onset
of commensurate SDW order at $T_{2}$ (Fig.~\ref{fig3}). Both SDW
ordering anomalies are superposed on the broad peak that is
centered near $\sim$40 K. Although we do not show the data here,
this broad peak is unchanged when magnetic fields as large as 9 T
are applied, indicating that it is not likely to be a Schottky
anomaly. Its insensitivity to fields suggests instead an
intrinsic electronic excitation of the incommensurate SDW with an
energy scale $\Delta \approx 0.45$ $k_{\rm B}T_{\rm N}$.

Electrical resistivity measurements can be used to estimate how
much of the Fermi surface in CaFe$_{4}$As$_{3}$ is gapped by the
incommensurate and commensurate SDW transitions. Analyses carried
out in other SDW systems have used high magnetic fields to
collapse the SDW gap, revealing the underlying background
resistivity $\rho_{\rm BG}$ of the gapless
state.~\cite{murayama,mazov} This approach is not possible in
CaFe$_{4}$As$_{3}$, where the small measured
magnetoresistance~\cite{zhao} implies that very large fields
would be required to close the SDW gap. Instead, we determine
$\rho_{\rm BG}$ in CaFe$_{4}$As$_{3}$ by a linear extrapolation of
the measured resistivity for $T\geq T_{\rm N}$, as shown in
Fig.~\ref{fig1}(b). We subtracted this estimate of $\rho_{\rm
BG}$ from  $\rho$, and normalized by $\rho$ to obtain
($\rho-\rho_{\rm BG})/\rho=\Delta\rho/\rho$, which represents the
percentage change in the resistivity that can be associated with
SDW formation. This is in turn proportional to the ratio of the
Fermi surface volumes in the gapped and ungapped
states.~\cite{murayama} The temperature dependence of
$\Delta\rho/\rho$ is plotted in Fig.~\ref{fig4}(b).
$\Delta\rho/\rho$ increases in an order parameter-like fashion
below $T_{\rm N}$, extrapolating to a value of $\approx$10\% as
$T\rightarrow0$ in the absence of the commensurate transition at
$T_{2}$. This result implies that the incommensurate SDW in
CaFe$_{4}$As$_{3}$ gaps about 10\% of the Fermi surface.

Our analysis of the specific heat is also consistent with a Fermi
surface in  CaFe$_{4}$As$_{3}$ that is only partially gapped by
SDW formation. We note that  $\gamma_{g}$ is slightly enhanced,
indicating that the quasiparticles that survive the Fermi surface
gapping at both $T_{\rm N}$ and $T_{2}$ have substantial
interactions that increase their effective masses and lead to the
heavy fermion behavior already noted for
CaFe$_{4}$As$_{3}$.~\cite{zhao} We point out that the $T=0$ value
of the quasiparticle gap associated with SDW formation, $\Delta$
is very similar to the 40 K electronic energy scale deduced from
the broad peak in the specific heat $C_{\rm p}-C_{\rm ph}$ above
$T_{2}$, suggesting a common origin.

Our measurements confirm the general scenario of Fermi surface
gapping that is accomplished by incommensurate SDW formation at
$T_{\rm N}$, followed by a second transition at $T_{2}$, where the
SDW becomes commensurate with respect to the underlying lattice.
The temperature dependence of the specific heat below $T_{2}$ is
well described within the BCS theory, and we present evidence for
a possible collective mode of the SDW gapped state, having a
characteristic energy of $\approx$ 0.25 $\Delta$.The overall
success of this analysis, which assumes an itinerant nature for
the magnetism in $\CFA$, is somewhat surprising, considering that
the neutron scattering measurements found a rather large Fe
moment in the paramagnetic state. Inelastic scattering
experiments are needed to assess the possibility that $\Delta$
instead represents an anisotropy gap in the transverse spin
excitations, suggestive of a more localized picture of the
magnetism.

We will combine these experimental results with electronic
structure calculations to elucidate the strength of the
correlations in $\CFA$, and to provide a phenomenological
understanding of the thermal and electrical properties described
above. The electronic structure of $\CFA$ was first considered in
its paramagnetic state.~\cite{todorov} It was subsequently
shown~\cite{nambu} that the large unit cell leads to a  much
larger Fermi surface with a different topology than those found
in the 1111 and 122 families of iron pnictides. Our analysis
shows that the strength of the correlations in CaFe$_4$As$_3$, at
least in the SDW phase, is substantially higher than in other
iron pnictides such as the 1111 and the 122 families, while being
comparable to the 111, and 2322 families, as well as the 11
family of iron chalcogenides.~\cite{imada, qimiaosi}  Our
calculations correctly reproduce the sign change in the density
of states at the Fermi level, predicted by the temperature
dependence of the thermopower. Finally, we ascribe the observed
low thermoelectric coefficient to a substantial particle hole
symmetry present near the Fermi level, and suggest compositional
modifications that may improve the thermoelectric performance of
$\CFA$.

An alternative interpretation of the thermoelectric power data
can be provided within a picture where all the electrons are
itinerant. In this view, all five Fe d bands are important in the
electronic structure of $\CFA$, and we assume that the
correlation strength is intermediate. By this we mean that the
correlations are not large enough to make the material
insulating, but also not so small that the wave function
renormalization $Z$ is less than 0.5. These initial assumptions
are consistent with previous calculations on
LaO$_{1-x}$F$_{x}$FeAs~\cite{Haule2008}, which found a mass
renormalization between 3 and 5 ($Z\sim0.2-0.3$), in good
agreement with subsequent optical experiments.~\cite{Basov2009}
Rather than calculating $Z$ from first principles for
CaFe$_4$As$_3$, we extract it from an analysis that combines the
results of the thermopower and specific heat measurements
described above with ab initio electronic structure calculations
carried out in both the paramagnetic and ordered states. The unit
cell needed to describe the incommensurate order for $T_{2}\leq
T\leq T_{\rm N}$ is prohibitively large, so we will restrict our
calculations to the paramagnetic state with $T\geq T_{\rm N}$ and
the commensurate SDW state with $T \leq T_{2}$.

We performed density functional theory (DFT) calculations in both
the low temperature SDW  and the high temperature paramagnetic
phases of $\CFA$. The calculation is done using  the projector
augmented wave (PAW) method~\cite{paw1, paw2} as implemented in
VASP~\cite{vasp} and the PBE exchange correlation
functional.~\cite{PBE} The lattice constants $a=11.852$ \AA,
$b=3.7352$ \AA, and $c=11.5490$ \AA~ as well as the atomic
coordinates are taken from the 15 K SDW state.~\cite{todorov} For
the paramagnetic state, we use a $8 \times 24 \times 8$ dense
mesh and a energy cutoff of 300 eV. To simulate the low
temperature SDW state with a Q=(0, 3/8 $\pi/b$, 0) wave
vector,~\cite{kanatzidis2010} we use a $1 \times 8 \times 1$
supercell of the PM unit cell and perform a non-collinear
magnetic calculation using a $6 \times 2 \times 6$ mesh and an
energy cutoff of 250 eV. The magnitude and arrangement of the
magnetic moments in our calculations agree well with those
deduced from experimental measurements.~\cite{kanatzidis2010}

\begin{figure}[bht]
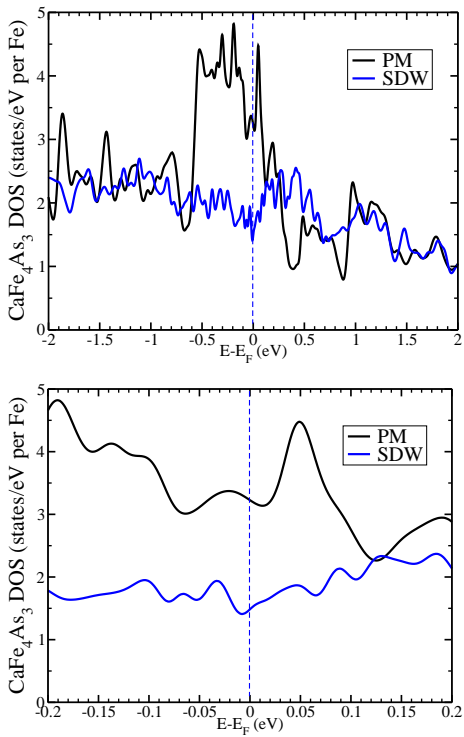

\includegraphics[width=0.7\linewidth]{Figure5_top.eps}
\vskip 2mm
\includegraphics[width=0.7\linewidth]{Figure5_bottom.eps}
\caption{(Color online) Upper panel: The density of states (DOS) for both the
high temperature paramagnetic (PM) phase and low temperature SDW
phase of CaFe$_4$As$_3$. An expanded view near the Fermi level is
presented in the lower panel.} \label{DOS}
\end{figure}

The density of states that results from these calculations in
both the high temperature paramagnetic state and in the low
temperature SDW state is presented in Fig. \ref{DOS}. We note with
interest that the energy derivative of the density of states
$N^{'}(\varepsilon)$ changes sign from negative in the
paramagnetic phase to positive in the SDW phase. The density of
states at the Fermi level is approximately twice as large in
ungapped and paramagnetic $\CFA$ than it is in SDW $\CFA$,
indicating that the SDW gap has removed a large fraction of the
high temperature Fermi surface.

It is our intention to combine experimental measurements of the
thermopower $S(T)$ with these DFT calculations of $N(\varepsilon)$
and its energy derivative $N^{'}(\varepsilon)$ to estimate the
quasiparticle renormalization factor $Z_{S}$. An independent
determination of $Z=Z_{\gamma}$ can be extracted from the
Sommerfeld coefficient $\gamma_{g}$, although this is only
possible in the SDW phase since the large phonon contribution to
the specific heat $C_{\rm p}$ prohibited an accurate determination
of $\gamma_{g}$ in the paramagnetic state. We obtain expressions
for the thermopower $S$ and Sommerfeld coefficient $\gamma_{g}$
with the aid of the corresponding local Fermi
equations,~\cite{Palsson} ignoring the asymmetry in the self
energy,~\cite{Haule-therm} and estimating the slope of the value
of a transport function $\Phi_{\varepsilon_{\rm F}}$ from the
derivative of the density of states calculated using DFT. $Z$ is
understood to be an average quasiparticle renormalization weight.
\begin{equation}
S=-\frac{k_B}{|e|} \frac{k_BT}{Z_{S}}
\frac{\Phi'(\varepsilon_{\rm F})}{\Phi(\varepsilon_{\rm F})}
\frac{E_2^1}{E_0^1}
\end{equation}
\begin{equation}
\gamma_{g}= \frac{\pi^2}{3} \frac{k_B^2}{Z_{\gamma}}
N(\varepsilon_{\rm F})
\end{equation}

where $k_B/|e|=8.6 \times 10^{-5}$ V/K, $N(\varepsilon_{\rm F})$
is the density of states at the Fermi energy $\varepsilon_{\rm
F}$, and $E_2^1=1.75$ and $E_0^1=0.82$.~\cite{Palsson} The
experimental input is listed in Table~\ref{mass}, and consists of
the experimental value of the SDW value of
$\gamma_{g}(T\rightarrow0)$ as well as two values for the
temperature derivative of $S$, one for the paramagnetic state,
which we approximate as $S(T_{\rm N})$ and one which represents
the gapped SDW state, which we set equal to $S(T_{2})$.  Both are
subsequently extrapolated to $T=0$, incurring substantial error
bars. We summarize the results of our analysis in
Table~\ref{mass}.

\begin{table}[htb]
\caption{Various quantities for CaFe$_4$As$_3$ in the SDW and PM
phases. Experimental values for the temperature derivative of the
thermopower d$S$/d$T$($\mu$V/K$^2$) extrapolated from $T=T_{\rm
N}$ (PM) and $T=T_{2}$ (SDW) to $T=0$, the calculated density of
states $N(\varepsilon_{\rm F})$ (states/eV/Fe) and its energy
derivative $N^{'}(\varepsilon_{\rm F})$ at the Fermi level
(states/eV$^2$-Fe), $\gamma_{g}$ is the measured Sommerfeld
coefficient $\gamma_{g}=(C_{\rm p}-C_{\rm ph})/T$ (J/Fe-mol
K$^{2}$), extrapolated to $T=0$. $Z_{\gamma}$ and $Z_{S}$ are the
calculated values of the quasiparticle renormalization determined
from measurements of the specific heat and thermopower,
respectively (see text).} \label{mass}
\begin{ruledtabular}
\begin{tabular}{ccccccc}
&$N(\varepsilon_{\rm F})$ &  $N^{'}(\varepsilon_{\rm F})$  &  d$S$/d$T$  &  $Z_S$  &$\gamma_{g}$ & $Z_{\gamma}$\\
\hline
SDW ($T\leq T_{2}$)  &  1.5       & 12(2)        & -0.8(3)       & 0.16(8) &0.03 &0.12 \\
PM  ($T\geq T_{\rm N}$)  &  3.2       & -6(2)        & 0.08(3)       & 0.37(20) & - & -  \\
\end{tabular}
\end{ruledtabular}
\end{table}

The values of $Z_{S}$ that are extracted from this analysis
indicate that $\CFA$ is very strongly correlated in the
paramagnetic state, and that the quasiparticle mass is more than
doubled with the onset of commensurate SDW order, an observation
that is echoed in the large Kadowaki-Woods ratio found in the low
temperature Fermi liquid phase. The values of $Z_{S}$ and
$Z_{\gamma}$  are in reasonable agreement in the SDW phase, and
support the conclusion that the SDW removes less correlated parts
of the Fermi surface, leaving quasiparticles that are much more
strongly correlated in the SDW phase of $\CFA$ than in other
members of the iron pnictide families. This is quite different
from what is found in other iron pnictide and chalcogenide
superconductors, in which the magnetic phases are more coherent
and less correlated than the paramagnetic phases.  The presence
of a large Sommerfeld coefficient implies that the SDW gap
$\Delta$ does not extend over the entire Fermi surface, although
the quasiparticles associated with the ungapped Fermi surface
appear to be strongly interacting at the lowest temperatures.

The large value of the Sommerfeld coefficient $\gamma_{g}$ in the
SDW phase suggests that $\CFA$ may have a large value of the
Seebeck coefficient via the Behnia-Flouquet ratio~\cite{behnia}
\begin{equation}
\frac{S}{\gamma_{g} T}=-\frac{3}{\pi^2|e|}
\frac{1}{N(\varepsilon_{\rm F})} \frac{\Phi'(\varepsilon_{\rm
F})}{\Phi(\varepsilon_{\rm F})} \frac{E_2^1}{E_0^1}~.
\end{equation}

However, we find instead that the thermoelectric power $S$ is
small at all temperatures, due to the weak energy dependence of
the density of states near the Fermi level. Although we find that
$\CFA$ has a rather small thermal conductivity $\kappa$ and as
well an electrical resistivity $\rho$ that is somewhat larger
than expected in a good metal, perhaps due to the strong
quasiparticle interactions implied by the small values of $Z$,
these factors are not sufficient to overcome the small values of
$S$, giving a disappointingly small value for the thermoelectric
figure of merit $\mathcal{Z}=S^{2}/\kappa\rho$ in $\CFA$. We find
that the largest values of $\mathcal{Z}T$ occur at 200 K, where
$\mathcal{Z}T=1.7\times10^{-4}$, indicating that similar
thermoelectric performances are found on high temperature range
in both CaFe$_{4}$As$_{3}$ and some skutterudites, such as
PrFe$_{4}$As$_{12}$.~\cite{sayles}

Clearly, the figure of merit $\mathcal{Z}$ in $\CFA$ is limited by
the relatively small thermopower, and Fig.~\ref{DOS} shows that
the reason is a surprising particle-hole symmetry that limits
$N^{'}(0)$. Provided that the strength of the electronic
correlations is not reduced, we believe that electron doping
should lead to a considerable enhancement of the thermopower. For
instance, the DFT density of states suggest that the thermopower
in the paramagnetic state changes sign and increases by about an
order of magnitude at the following doping levels:  9-14 $\%$ on
the Fe site, 12-19 $\%$ on the As site, and 35-50 $\%$ on the Ca
site. However, if many-body renormalizations are included, the
required doping levels can be expected to be substantially
reduced from the levels predicted by DFT. Similarly, we speculate
that superconductivity might be induced in $\CFA$ by improving
its metallic character, perhaps by a dopant that reduces the
lattice constant. Further experimental and theoretical work is
needed to establish whether a localized or itinerant picture is
more appropriate for $\CFA$.

\begin{acknowledgments}
Work at Brookhaven National Laboratory was carried out under the
auspices of the U.S. Department of Energy, Office of Basic Energy
Sciences under Contract No. DE-AC02-98CH1886. Work at Rice
University and at Rutgers (GK) is supported by DoD MURI "Towards
New and Better High Temperature Superconductors". Work at Rutgers
(ZY) was carried out under the auspices of a DoD National
Security Science and Engineering Faculty Fellowship, via AFOSR
grant FA 9550-10-1-0191.
\end{acknowledgments}


\end{document}